# HIERARCHICAL MULTI-SCALE GRAPH LEARNING WITH KNOWLEDGE-GUIDED ATTENTION FOR WHOLE-SLIDE IMAGE SURVIVAL ANALYSIS

*Bin Xu[1#], Yufei Zhou[2#], Bolin Song[3], Jingwen Sun[4], Yang Bian[1], Yikai Chen[1], Cheng Lu[5], Ye Wu[6], Jianfei Tu[7], Xiangxue Wang[1*]*

1. Nanjing University of Information Science and Technology, Nanjing, China
2. Case Western Reserve University, Cleveland, OH, USA
3. Georgia Institute of Technology and Emory University, Atlanta, GA, USA
4. Nanjing Women and Children's Healthcare Hospital, Nanjing, China
5. Guangdong Provincial People's Hospital, Guangzhou, China
6. Nanjing University of Science and Technology, Nanjing, China
7. Lishui Hospital of Zhejiang University, Lishui, China
# These authors contribute equally *Corresponding author

**ABSTRACT**

We propose a Hierarchical Multi-scale Knowledge-aware Graph Network (HMKGN) that models multi-scale interactions and spatially hierarchical relationships within whole-slide images (WSIs) for cancer prognostication. Unlike conventional attention-based MIL, which ignores spatial organization, or graph-based MIL, which relies on static handcrafted graphs, HMKGN enforces a hierarchical structure with spatial locality constraints, wherein local cellular-level dynamic graphs aggregate spatially proximate patches within each region of interest (ROI) and a global slide-level dynamic graph integrates ROI-level features into WSI-level representations. Moreover, multi-scale integration at the ROI level combines coarse contextual features from broader views with fine-grained structural representations from local patch-graph aggregation. We evaluate HMKGN on four TCGA cohorts (KIRC, LGG, PAAD, and STAD; N=513, 487, 138, and 370) for survival prediction. It consistently outperforms existing MIL-based models, yielding improved concordance indices (10.85% better) and statistically significant stratification of patient survival risk (log-rank $p < 0.05$).

## 1. INTRODUCTION

The rapid development of digital pathology has made whole-slide images (WSIs) a crucial foundation for computer-assisted tumor diagnosis, subtyping, and prognosis assessment[1]. WSIs contain rich morphological information across scales, simultaneously capturing cellular-level morphology and slide-level architecture, which are inherently complementary for survival analysis[2]. Accurate survival prediction plays a crucial role in clinical decision-making by guiding personalized treatment planning, assessing recurrence risk, and predicting patient prognosis[3].

There is a recent surge of multi-instance learning (MIL) in computational pathology. MIL-based approaches have achieved remarkable progress in slide-level classification and survival prediction, as they enable the end-to-end WSI-level modeling by learning the interaction and aggregation of individual patches across tissue volumes [4]. Many attention-based MIL approaches, such as TransMIL, are capable of effectively aggregating an arbitrary number of patches into a WSI-level representation [5], [6]. However, these methods treat individual patches as spatially irrelevant instances, neglecting the inherent spatial arrangement and morphological continuity across nearby patches. As a result, the derived slide-level representations may overlook key spatial context and structural dependencies that are critical for capturing tissue organization and pathological patterns. Meanwhile, graph-based MIL explicitly encodes spatial relationships among patches using graph structures. Yet, most existing methods depend on static, handcrafted graphs defined purely by spatial proximity, which restricts their ability to model broader contextual relationships [7]. Recent advances, such as the knowledge-guided WiKG framework, on the other hand, have demonstrated the benefit of dynamically learning contextual dependencies between patches[8]. Nevertheless, WiKG remains limited by the absence of spatial locality constraints, which permits interactions between distantly located regions that are less likely to be biologically related. Moreover, it operates on single-scale data, which restricts its ability to capture hierarchical and multi-scale contextual information, a limitation in diseases characterized by spatial heterogeneity (e.g., pancreatic cancers [9]).

To address these limitations, we propose a Hierarchical Multi-scale Knowledge-aware Graph Network (HMKGN) for WSI-based survival analysis. Our core innovations are: (1) spatial locality-constrained hierarchical learning, which aggregates spatially proximate patches into ROI-level representations and hierarchically integrates them into slide-level features to reflect tissue organization. (2) multi-scale feature fusion, which combines coarse contextual representations from lower-magnification views with fine-grained morphological information from high-magnification patches, enabling comprehensive modeling of both regional tissue context and cellular-level structure [10].

## 2. METHOD

### 2.1. Data Preparation

As shown in Fig. 1-a, the tissue regions in WSIs were first divided into non-overlapping patches of size $224 \times 224$ at low magnification (5×). Each 5× patch $P_{\text{low}}^{(i,j)}$ was spatially aligned with its corresponding high-magnification region $R_{\text{high}}^{(i,j)} \in \mathbb{R}^{896 \times 896}$, which was further subdivided into a $4 \times 4$ grid of 20× patches $\{P_{\text{high}}^{(i,j,k)}\}_{k=1}^{16}$. Specifically, $i \in \{1, \ldots, N\}$ indexes the WSIs in the dataset, $j \in \{1, \ldots, n_i\}$ indexes the low-magnification patches within the $i$-th WSI, and $k \in \{1, \ldots, 16\}$ indexes the high-magnification patches corresponding to each low-magnification patch.

Both low- and high-magnification patches were then embedded into feature vectors using a resolution-adaptive foundation model (UNIv2 [11]):

$$f_{\text{low}}^{(i,j)} = \Phi_{\text{UNI}}\left(P_{\text{low}}^{(i,j)}\right), f_{\text{high}}^{(i,j,k)} = \Phi_{\text{UNI}}(P_{\text{high}}^{(i,j,k)}), \quad (1)$$

where $\Phi_{\text{UNI}}(\cdot)$ denotes the encoder of the frozen UNIv2 model.

This explicit multi-scale correspondence ensures pixel-level alignment between $f_{\text{low}}^{(i,j)}$ and $\{f_{\text{high}}^{(i,j,k)}\}$, providing a geometric constraint for subsequent multi-scale feature integration.

### 2.2. Spatial Locality-Constrained Hierarchical Learning

To capture the hierarchical structure of high-magnification patches with spatial locality constraints, HMKGN comprises two stages: (1) local aggregation (patch-to-ROI) and (2) global aggregation (ROI-to-WSI).

During the local aggregation stage (Fig. 1-b1), a spatial locality constraint was applied such that only patches within a $4 \times 4$ grid were directly connected through a local dynamic graph $\mathcal{G}_{dyn}^{local}$. For each high magnification patch node feature $\{f_{\text{high}}^{(i,j,k)}\}$ in $\mathcal{G}_{dyn}^{local}$, its fine-grained ROI-level representation was obtained via:

$$f_{\text{high}}^{(i,j)} = \text{Agg}(\{f_{\text{high}}^{(i,j,k)}\}_{k=1}^{16}), \quad (2)$$

which captures fine-grained cellular and morphological semantics within spatially coherent local regions. Each $f_{\text{high}}^{(i,j)}$ representation was then fused with the coarse contextual representation $f_{\text{low}}^{(i,j)}$ from its spatially aligned region, yielding the ROI-level embedding $f_{\text{ROI}}^{(i,j)}$ (section 2.3).

During the global aggregation stage (Fig. 1-b2), all ROI-level embeddings $\{f_{\text{ROI}}^{(i,j)}\}$ from local dynamic graphs ($\mathcal{G}_{dyn}^{local}$) were integrated through a global dynamic graph $\mathcal{G}_{dyn}^{global}$ to derive the slide-level representation:

$$f_{\text{WSI}}^{(i)} = \text{Agg}(\{f_{\text{ROI}}^{(i,j)}\}), \quad (3)$$

This hierarchical message passing jointly models local dependencies and global tissue context. The aggregation function Agg here corresponds to the knowledge-aware dynamic graph network (KGN) described in Section 2.4, which we build upon and extend with spatial locality constraints and multi-scale fusion.

### 2.3. Multi-scale Feature Fusion

As shown in Fig. 1-c, for each ROI, coarse contextual representations $f_{\text{low}}^{(i,j)}$ derived from lower-resolution ROI images were fused with fine-grained features $f_{\text{high}}^{(i,j)}$ extracted from high-resolution patch graphs through a bidirectional cross-attention mechanism (BiX) [12]. Specifically, the queries (Q), keys (K), and values (V) were defined as

$$\begin{aligned} Q_L &= W_Q f_{\text{low}}^{(i,j)}, K_H = W_K f_{\text{high}}^{(i,j)}, V_H = W_V f_{\text{high}}^{(i,j)}, \\ Q_H &= W_Q f_{\text{high}}^{(i,j)}, K_L = W_K f_{\text{low}}^{(i,j)}, V_L = W_V f_{\text{low}}^{(i,j)}. \end{aligned} \quad (4)$$

The cross-attention in both directions was computed as

$$\begin{aligned} \text{Attn}_{L \to H} &= \text{Softmax}(\frac{Q_L K_H^{\mathsf{T}}}{\sqrt{d}}) V_H, \\ \text{Attn}_{H \to L} &= \text{Softmax}(\frac{Q_H K_L^{\mathsf{T}}}{\sqrt{d}}) V_L, \end{aligned} \quad (5)$$

and the fused regional representation was obtained by concatenating the two attention outputs:

$$f_{\text{ROI}}^{(i,j)} = [\,\text{Attn}_{L \to H};\, \text{Attn}_{H \to L}\,]. \quad (6)$$

This multi-scale integration enables the model to jointly capture both coarse regional tissue contextual semantics (low magnification) and fine-grained cellular-level morphology (high magnification) within a unified ROI representation.

### 2.4. Knowledge-aware attention-based Dynamic Graph

All patch-level features $\{f_{\text{high}}^{(i,j,k)}\}$ are treated as nodes in the ROI-level patch graph $\mathcal{G}_{dyn}^{local}$ during the local aggregation. All ROI-level features $\{f_{\text{ROI}}^{(i,j)}\}$ (with multi-scale feature fusion) are treated as nodes in the slide-level graph $\mathcal{G}_{dyn}^{global}$ during the global aggregation. Message passing across this graph captures long-range structural dependencies and integrates regional information into a global slide representation $f_{\text{WSI}}^{(i)}$. The final WSI-level feature vector $f_{\text{WSI}}^{(i)}$ provides a coherent multi-scale embedding suitable for downstream tasks such as classification, molecular subtyping, and survival prediction. While the ROI-level graph module is inspired by knowledge-aware graph modeling techniques (e.g., WiKG), our framework extends this foundation by introducing the spatial locality constrained hierarchical structure and multi-scale feature fusion, enabling explicit modeling of multi-scale interactions and improving representation quality for WSI analysis.

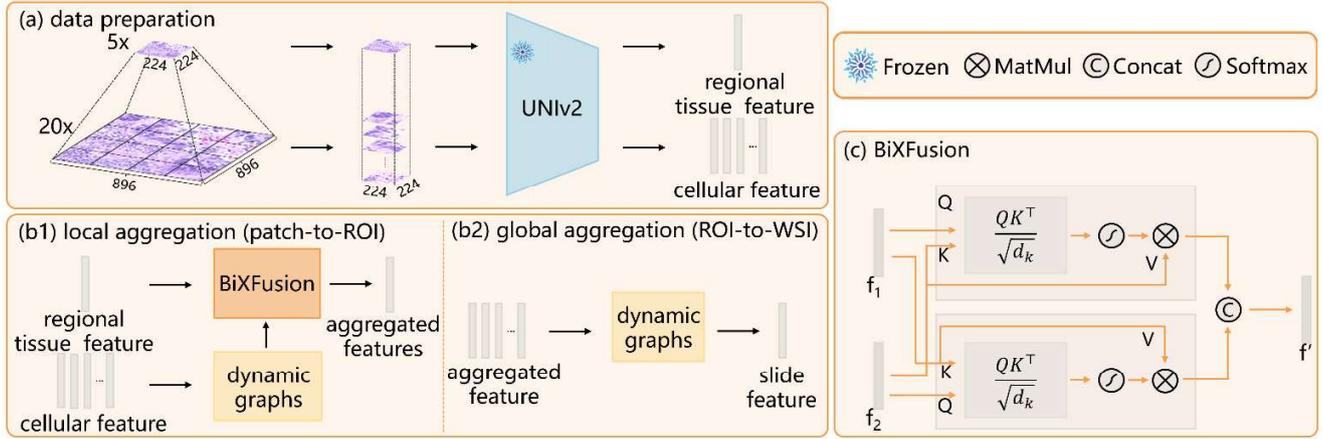

**Fig. 1. Framework of the Proposed HMKGN Method**.

### 2.5. Slide-level Survival Prediction with Hierarchical Modeling and Multi-scale Fusion

The output aggregated slide-level features are used to predict patients' survival end-to-end with a discrete-time negative log-likelihood survival loss (NLLSurvLoss). Importantly, slide-level predictions are informed by the hierarchical multi-scale graph, where fine-grained cellular-level features and coarse regional tissue contextual semantics features are aggregated under spatial locality constraints. The BiX effectively integrates the fine-grained high-magnification information with coarse low-magnification context, ensuring that both local morphology and global tissue architecture contribute to survival modeling. This multi-scale fusion preserves cross-level semantic consistency and enhances the model's ability to capture both short- and long-range prognostic dependencies, providing a robust foundation for WSI-based survival prediction.

### 3. RESULT AND DISCUSSION

We evaluated the proposed HMKGN against six established multiple-instance learning (MIL) and graph neural network (GNN)-based methods, including MaxMIL, MeanMIL, DAttention[5], TransMIL[6], MambaMIL[13], and WiKG[8], on four TCGA datasets (KIRC, LGG, PAAD, and STAD).

Table 1 summarizes the concordance index (C-index) of each model, reported as mean ± standard deviation (SD) based on 4-fold cross-validation. HMKGN consistently outperforms existing methods across all cancer types and achieves 4.20% to 33.56% higher C-indices on the four datasets. These results demonstrate the efficacy of integrating multi-scale hierarchical features and BiX with knowledge-aware graph attention for WSI survival analysis.

**Table 1. Performance comparison of HMKGN and baseline models on TCGA datasets (C-index ± SD). (*) denotes significant logrank p-values.**

| Model | KIRC | LGG | PAAD | STAD | Mean Improve(%) |
|---|---|---|---|---|---|
| MaxMIL | 0.7273 ± 0.0390(*) | 0.6148 ± 0.0377(*) | 0.5743 ±0.0730 | 0.5293 ± 0.0493(*) | 16.50 |
| MeanMIL | 0.7131 ± 0.0253(*) | 0.7549 ± 0.0509(*) | 0.6095 ± 0.0744(*) | 0.6152 ± 0.0908(*) | 5.82 |
| DAttention | 0.7266 ± 0.0282(*) | 0.7578 ± 0.0457(*) | 0.5949 ± 0.0803 | 0.6191 ± 0.0869(*) | 5.59 |
| TransMIL | 0.5241 ± 0.0563 | 0.5685 ± 0.0450 | 0.4828 ± 0.0533 | 0.5579 ± 0.0336 | 33.56 |
| MambaMIL | 0.7192 ± 0.0254(*) | 0.7623 ± 0.0577(*) | 0.6146 ± 0.0382(*) | 0.6384 ± 0.0565(*) | 4.20 |
| WiKG | 0.7176 ± 0.0379(*) | 0.7648 ± 0.0449(*) | 0.6185 ± 0.0690(*) | 0.6170 ± 0.0552(*) | 4.83 |
| **HMKGN (Our Model)** | **0.7486 ± 0.0298(*)** | **0.7699 ± 0.0522(*)** | **0.6897 ± 0.0723(*)** | **0.6411 ± 0.0565(*)** | |

As shown in Table 2, to evaluate the contribution of each component, we performed ablation studies on three key modules—hierarchical modeling, multi-scale interaction, and spatial locality constraint.

**Table 2. Ablation study evaluating the effects of hierarchical modeling, multi-scale interaction, and spatial locality constraint across TCGA cohorts. (*) denotes significant logrank p-values.**

| Model | Hierarch. | Locality | Multi-scale | KIRC | LGG | PAAD | STAD |
|---|---|---|---|---|---|---|---|
| KGN (WiKG) | no | N/A | no | 0.7176 ± 0.0379(*) | 0.7648 ± 0.0449(*) | 0.6185 ± 0.0690(*) | 0.6170 ± 0.0552(*) |

| Model | Hierarch. | Locality | Multi-scale | KIRC | LGG | PAAD | STAD |
|---|---|---|---|---|---|---|---|
| HMKGN$_{\text{single-scale}}$ | yes | yes | no (20x only) | 0.7446 ± 0.0217(*) | 0.7657 ± 0.0456(*) | 0.6327 ± 0.0404(*) | 0.6390 ± 0.0741(*) |
| HMKGN$_{\text{no\_locality}}$ | yes | no | N/A | 0.7377± 0.0316 (*) | 0.7584± 0.0309(*) | 0.6341± 0.0611(*) | 0.5984± 0.0627(*) |
| HMKGN | yes | yes | yes | 0.7486 ± 0.0298(*) | 0.7699 ± 0.0522(*) | 0.6897 ± 0.0723(*) | 0.6411 ± 0.0565(*) |

With only the hierarchical modeling, HMKGN$_{\text{single-scale}}$ solely utilizes $\{f_{\text{high}}^{(i,j)}\}$ to represent $\{f_{\text{ROI}}^{(i,j)}\}$, where $f_{\text{high}}^{(i,j)}$ is obtained by aggregating high-resolution patch features $f_{\text{high}}^{(i,j,k)}$ within each ROI region, without incorporating the low-resolution patch features $f_{\text{low}}^{(i,j)}$. Despite operating on a single scale, HMKGN$_{\text{single-scale}}$ outperforms the KGN baseline across KIRC, PAAD, and STAD (+2.2% to 3.6% on average), while performing comparably in LGG. This confirms that explicitly modeling region-level spatial hierarchy enables the network to aggregate semantically related patches and capture broader tissue context.

With both hierarchical modeling and bidirectional multi-scale fusion present, HMKGN further combines the low-resolution features $f_{\text{low}}^{(i,j)}$ with the high-resolution aggregated representation $f_{\text{high}}^{(i,j)}$ and outperforms HMKGN$_{\text{single-scale}}$ (by 0.3% to 9%), with the largest gain observed in PAAD. This likely reflects the pronounced spatial heterogeneity of pancreatic adenocarcinoma in PAAD [9], which can be captured by multi-scale learning through integration of fine-grained morphological details and coarse contextual structures.

When the spatial locality constraint was removed (HMKGN$_{\text{no\_locality}}$), the model exhibited a consistent decline (-1.5% to -8.8%) across all cohorts compared to HMKGN, especially in PAAD (-8.8%) and STAD (-7.1%). Therefore, preserving local spatial order among high-magnification patches is essential for capturing prognostically relevant micro-architectural patterns.

Overall, our approach (HMKGN) achieves the best C-index and statistically significant log-rank p-values ($p < 0.05$). This demonstrates that the combination of hierarchical structure, reciprocal multi-scale fusion, and spatial locality constraint leads to a more discriminative and biologically meaningful multi-scale representation for survival prediction performance across all TCGA cohorts.

## 4. CONCLUSION

In this study, we introduced HMKGN, a hierarchical graph framework that unifies spatial locality constraints, hierarchical aggregation, and feature fusion to enhance multi-scale representation learning across different magnifications in WSIs. The proposed model achieved superior survival prediction performance across four TCGA datasets with over 1500 patients. These results highlight the strong potential of HMKGN for clinical translation, particularly in improving prognostic assessment. In future work, we plan to extend HMKGN to multi-modal survival analysis by integrating pathological, radiological, and clinical data.

## 5. ACKNOWLEDGMENTS

This work is made possible by the National Natural Science Foundation of China under award numbers 62301265. The authors have no relevant financial or non-financial interests to disclose.

## 6. REFERENCES

[1] G. Litjens et al., "A survey on deep learning in medical image analysis," Med. Image Anal., vol. 42, pp. 60–88, 2017.

[2] R. J. Chen et al., "Scaling vision transformers to gigapixel images via hierarchical self-supervised learning," in Proceedings of the IEEE/CVF conference on computer vision and pattern recognition, 2022, pp. 16144–16155.

[3] K. Bera, K. A. Schalper, D. L. Rimm, V. Velcheti, and A. Madabhushi, "Artificial intelligence in digital pathology—new tools for diagnosis and precision oncology," Nat. Rev. Clin. Oncol., vol. 16, no. 11, pp. 703–715, 2019.

[4] P. Sandarenu et al., "Survival prediction in triple negative breast cancer using multiple instance learning of histopathological images," Sci. Rep., vol. 12, no. 1, p. 14527, 2022.

[5] M. Ilse, J. Tomczak, and M. Welling, "Attention-based deep multiple instance learning," in International conference on machine learning, PMLR, 2018, pp. 2127–2136.

[6] Z. Shao et al., "Transmil: Transformer based correlated multiple instance learning for whole slide image classification," Adv. Neural Inf. Process. Syst., vol. 34, pp. 2136–2147, 2021.

[7] S. Javed et al., "Cellular community detection for tissue phenotyping in colorectal cancer histology images," Med. Image Anal., vol. 63, p. 101696, 2020.

[8] J. Li et al., "Dynamic graph representation with knowledge-aware attention for histopathology whole slide image analysis," in Proceedings of the IEEE/CVF conference on computer vision and pattern recognition, 2024, pp. 11323–11332.

[9] D. Cui Zhou et al., "Spatially restricted drivers and transitional cell populations cooperate with the microenvironment in untreated and chemo-resistant pancreatic cancer," Nat. Genet., vol. 54, no. 9, pp. 1390–1405, Sep. 2022, doi: 10.1038/s41588-022-01157-1.

[10] S. Diao et al., "Deep multi-magnification similarity learning for histopathological image classification," IEEE J. Biomed. Health Inform., vol. 27, no. 3, pp. 1535–1545, 2023.

[11] R. J. Chen et al., "Towards a General-Purpose Foundation Model for Computational Pathology," Nat. Med., 2024.

[12] X. Wang, P. Guo, and Y. Zhang, "Domain adaptation via bidirectional cross-attention transformer," ArXiv Prepr. ArXiv220105887, 2022.

[13] S. Yang, Y. Wang, and H. Chen, "Mambamil: Enhancing long sequence modeling with sequence reordering in computational pathology," in International conference on medical image computing and computer-assisted intervention, Springer, 2024, pp. 296–306.